\documentclass[12 pt]{article}

\title{A Numerical Simulation of the Reconnection Layer in
2D Resistive MHD}
\author{D.~A.~Uzdensky\thanks{Currently at the University of Chicago.} 
and R.~M.~Kulsrud\\
\it{Princeton Plasma Physics Laboratory, P.O.Box 451}\\
\it{Princeton University, Princeton, NJ 08543}}
\date{March 21, 2000}

\begin{document}
\input psfig.sty

\maketitle
\begin{abstract}

In this paper we present a two-dimensional, time dependent,
numerical simulation of a reconnection current layer in 
incompressible resistive magnetohydrodynamics with uniform 
resistivity in the limit of very large Lundquist numbers. 
We use realistic boundary conditions derived consistently 
from the outside magnetic field, and we also take into 
account the effect of the back pressure from flow into 
the the separatrix region. We find that within a few 
Alfv{\'e}n times the system evolves from an arbitrary 
initial state to  a steady state consistent with the 
Sweet--Parker model, even if the initial state is 
Petschek-like.

~\\
\noindent PACS Numbers: 52.30.Jb, 96.60.Rd, 47.15.Cb.
\end{abstract}

\section{Introduction}
\label{sec-intro}

Magnetic reconnection is of great interest in many space and 
laboratory plasmas \cite{Kulsrud-1998,MRX-Yamada}, and has been 
studied extensively for more than four decades. The most important 
question is that of the reconnection rate. The process of magnetic 
reconnection, is so complex, however, that this question is still 
not completely resolved,  even within the simplest possible {\it 
canonical} model: two-dimensional (2D) incompressible resistive 
magnetohydrodynamics (MHD) with uniform resistivity $\eta$ in the 
limit of $S\rightarrow \infty$ (where $S=V_A L/\eta$ is the global 
Lundquist number, $L$ being the half-length of the reconnection layer). 
Historically, there were two drastically different estimates for the 
reconnection rate: the Sweet--Parker model \cite{Sweet-1958,Parker-1963} 
gave a rather slow reconnection rate ($E_{\rm SP} \sim S^{-1/2}$), 
while the Petschek \cite{Petschek-1964} model gave any reconnection 
rate in the range from $E_{\rm SP}$ up to the fast maximum Petschek 
rate $E_{\rm Petschek} \sim 1/\log S$. Up until the present it was 
still unclear whether Petschek-like reconnection faster than Sweet--Parker 
reconnection is possible. Biskamp's simulations \cite{Biskamp-1986} are 
very persuasive that, in resistive MHD, the rate is generally that of 
Sweet--Parker. Still, his simulations are for $S$ in the range of a 
few thousand, and his boundary conditions are somewhat tailored to 
the reconnection rate he desires, the strength of the field and the 
length of layer adjusting to yield the Sweet--Parker rate. Thus, a 
more systematic boundary layer analysis is desirable to really settle
the question. In particular, one needs an elaborate and detailed picture 
of the reconnection current layer --- namely, a picture that features a 
realistic model for the variation of the outside magnetic field along the 
layer, and realistic 2D profiles of the plasma parameters inside the layer. 

\begin{figure} [tbp]
\centerline {\psfig{file=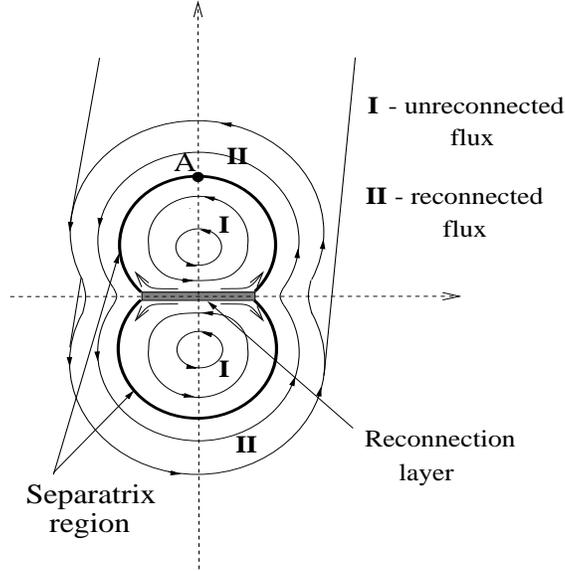,height=3 in,width=3 in,angle=90}}
\caption[The global geometry]
{The global geometry.}
\label{fig-global}
\end{figure}

The development of such a framework is the main goal of the present
paper. We believe that the methods developed in this paper are rather 
universal and can be applied to a very broad class of reconnecting 
systems that include more realistic physics. However, for definiteness 
and clarity we keep in mind a particular global geometry, that presented 
in Fig.~\ref{fig-global} (although we do not use it explicitly in our 
present analysis). This Figure shows the situation somewhere in the 
middle of the process of merging of two plasma cylinders. Regions~I 
and II are ideal MHD regions: regions~I represent unreconnected flux,
and region~II represents reconnected flux. The two regions~I are 
separated by the very narrow {\it reconnection current layer}. Plasma 
from regions~I enters the reconnection layer and gets accelerated along 
the layer, finally entering the {\it separatrix region}, also between 
regions~I and II. In general, both the reconnection layer and the 
separatrix region require resistive treatment. 

The plasma entering the separatrix from the reconnection layer 
is traveling at nearly the Alfv{\'e}n speed. It crashes into 
the plasma at rest on the separatrix lines, that has not passed 
through the reconnection layer, but which got there by direct
${\bf E \times B}$ motion across the lines as the position of 
the separatrix changed by reconnection. This crash generates 
considerable heat, and hence back pressure on the reconnection 
layer. However, the separatrix region is continually in transition 
since different plasma occupies it as the reconnection proceeds so 
that the heated plasma is moved into the reconnected region. Thus, 
the plasma encountered by the outflowing reconnected plasma is 
continually refreshed and can always be taken initially at rest.  
Therefore, there is a time delay before the back reaction sets in.
In our paper, we attempt to model this dynamical  behavior of the 
separatrix plasma as accurately as possible.

In the limit $S\rightarrow \infty$ the reconnection rate is slow compared
with the Alfv{\'e}n time $\tau_A=L/V_A$, which allows one to break the whole 
problem into the global problem and the local problem. In a previous paper 
Ref.~\cite{Uzdensky-1996}, we argued  that on the {\it global scale} (i.e., 
on the scale of order the half-length of the layer~$L$) the time evolution 
of the reconnecting system can be described as a sequence of magnetostatic 
equilibria. In  paper~\cite{Uzdensky-1997} we explained that the role of 
the global solution is to give the general geometry of the reconnecting 
system, the position and the length of the reconnection layer and of the 
separatrix, and the boundary conditions for the local problem (which, in 
turn, determines the reconnection rate). These boundary conditions are 
expressed in terms of the outside magnetic field~$B_{y,0}(y)$, where $y$ 
is the direction along the layer. In particular, $B_{y,0}(y)$ provides the 
characteristic global scales: the half-length of the layer~$L$, defined 
as the point where $B_{y,0}(y)$ has minimum, and the global Alfv{\'e}n 
speed, defined as $V_A=B_{y,0}(0)/\sqrt{4\pi\rho}$. It is important to 
understand that the global solution is essentially independent of the 
local reconnection physics.

In this paper we study the local problem concerning the reconnection 
layer itself. Our main goal here is to determine the internal structure 
of a steady state reconnection current layer (i.e., to find the 2D~profiles 
of plasma velocity and magnetic field), and the reconnection rate represented 
by the (uniform) electric field~$E$. Our most important result is that, in the
case of uniform resistivity, there is a unique stable steady-state solution,
and this is essentially the Sweet--Parker solution. We show that Petschek's 
solution (which is not unique and supposedly encompasses a variety of 
reconnection solutions from that of Sweet and Parker to a solution 
reconnecting at almost the Alfv{\'e}n speed) actually relaxes to the 
single Sweet-Parker solution.

First, in Section~\ref{sec-model} we discuss the assumptions of our physical 
model of the layer in some detail. Then, in Section~\ref{sec-rescaled} we 
introduce the rescaled equations representing the mathematical model of our 
problem. In Section~\ref{sec-simulations} we present our numerical simulations.
And, finally, in Section~\ref{sec-conclusions} we give our conclusions.

\section{Physical Model}
\label{sec-model}

By the local problem we mean the analysis of the \emph{internal 
structure} of the reconnection layer and the separatrix layer, 
which is necessary for the determination of the reconnection 
rate. Since in ideal MHD these layers are current sheets of 
zero thickness, resolving their inner structure cannot be 
done in this ideal framework and requires the addition of 
some new nonideal physics. 

Historically, most important, and conceptually most interesting for 
the purpose of resolving the current layer, is the inclusion of small 
resistivity. The model in which the only nonideal effect is that 
of the resistivity appears to be the simplest model with the minimal 
required complexity needed to resolve the current density singularity 
of an ideal MHD solution.

Thus, in this analysis we \emph{assume} that the only new nonideal 
physical process is small constant and uniform resistivity $\eta$ 
(and perhaps  viscosity~$\nu$; see discussion below).

The inclusion of the small resistivity means the introduction
of a new dimensionless small parameter associated with it, 
namely the inverse Lundquist number $S^{-1}=\eta/L V_A\ll 
1$, which is considered the \emph{primary small parameter} 
of the problem. This means that we are interested in studying
the case of $S\rightarrow \infty$, and we want to determine  how 
the parameters of the layer, such as its thickness and the 
reconnection rate, scale with $S$ in the leading order. As 
we shall see in the next section, making use of this small 
parameter helps to simplify the problem significantly while 
keeping all essential features, such as the two-dimensional 
nature of the problem, intact.

The next assumption we make concerns a steady state. The \emph{steady-state} 
condition means that parameters of the current layer as well as the boundary 
conditions change very slowly compared with the global Alfv{\'e}n time, which 
is the characteristic time spent by a fluid element inside the reconnection 
layer. A very important consequence of the steady-state condition in 2D 
geometry is that, due to the Maxwell equation $\nabla \times {\bf E} =- 
(1/c)\ \partial{\bf B}/ \partial t$, the $z$~component of the electric 
field is uniform: $E_z(x,y)=E={\rm const}$. 

As for the plasma viscosity, it does not seem to be necessary to 
include it, because viscosity, unlike resistivity, does not play 
any role in the actual breaking the lines of force. However, we
always include a small constant and uniform viscosity~$\nu$ for 
two reasons. First, in our numerical simulations we include it for 
numerical stability. The second and more important reason is that 
the consideration of very small (even compared with the resistivity) 
viscosity is useful to correctly understand some important features 
of the magnetic field configuration at the very center of the current 
layer (see Ref.~\cite{Uzdensky-1998}).

Next, most of the classical models of reconnection, including 
both Sweet--Parker and Petschek, assume that the outside merging 
magnetic field is uniform. But this assumption  actually prohibits 
one from formulating the problem in a mathematically complete and 
consistent manner, because the downstream boundary conditions for 
the flow cannot be correctly specified. In this quasi-one-dimensional 
framework, there is no natural end of the layer; in particular, there 
is no way to define the global scale~$L$. This, in turn, makes all 
attempts to get some scalings for the reconnection rate with the 
Lundquist number essentially meaningless, since the definition of 
the Lundquist number involves~$L$.

Now, in our paper a generic and more or less representative variation 
of the outside magnetic field $B_{y,0}(y)$ along the layer and along the 
separatrix [where it is called $B_{s,1}(l)$] is included as an integral 
part of the problem. In particular, the global scale (the half-width $L$ 
of the layer) is defined naturally as the distance along the midplane from 
the center of the current layer to one of the two endpoints --- i.e., the 
points where the outside magnetic field goes through a minimum and where 
the separatrices branch off the midplane $x=0$. In fact, this global 
scale is the characteristic scale for the function $B_{y,0}(y)$. 
Since this function is determined by the global ideal MHD solution 
(see Ref.~\cite{Uzdensky-1997}), the scale~$L$ is, by definition, 
independent of the physics of the resistive layer. 

Thus, the nonuniformity of the outside magnetic field along the
layer makes the problem essentially \emph{two-dimensional} (rather
than one-dimensional). One practically important consequence of 
this fact is that the problem becomes much more complicated 
mathematically, so that one has to abandon any hope for a nice 
analytical solution and to resort to numerical simulation instead.

Thus, the physical model of the reconnection layer that we are going to 
use in this paper for treating the local problem, can be summarized as
\begin{quote}
{\it two-dimensional, steady-state, incompressible, resistive 
MHD with constant and uniform resistivity (and perhaps viscosity)
in the limit of very large Lundquist number.}
\end{quote}

Perfect mirror symmetry is assumed with respect to both the $x$~axis
and the $y$~axis.%
\footnote
{Due to this symmetry, $v_y$ and $B_x$ are even and $v_x$ and $B_y$ 
are odd with respect to $x=0$, and $v_x$ and $B_y$ are even and $v_y$ 
and $B_x$ are odd with respect to $y=0$. Thus, $v_x=0$ along $x=0$ and 
$v_y=0$ along $y=0$, which means that the two axes of symmetry are stream 
lines.}

We call this model the \emph{canonical reconnection layer model}.

When considering the local problem in this model, we use the global 
scale in the $y$~direction (along the layer), and the local scale 
in the $x$~direction (across the layer). The outside magnetic field 
$B_{y,0}(y)$ determined by the global solution here plays the role 
of a boundary condition at $x\rightarrow \infty$.

Finally, although in this section we talked only about the 
reconnection layer itself, the same physical model applies 
also to the separatrix layer.

\section{System of Rescaled Equations}
\label{sec-rescaled}

In accordance with our physical mode, 
we can now write down the set of two-dimensional steady-state fluid 
equations for our system. These equations are:

(i) The incompressibility condition:
$$ \nabla \cdot {\bf v} = {{\partial v_x}\over{\partial x}}+
{{\partial v_y}\over{\partial y}} = 0.                       \eqno{(1)} $$

(ii) The $z$~component of Ohm's law:
$$ \eta j_z = E + v_x B_y - v_y B_x,                         \eqno{(2)} $$
where $E \equiv E_z = {\rm const}$.

(iii) The equation of motion (with the viscosity):
$$  {\bf v} \cdot \nabla {\bf v} = - \nabla p +[{\bf j_z} \times {\bf B}]
+ \nu \nabla^2  {\bf v},                                     \eqno{(3)} $$
where the density is set to one.

Now we take the crucial step in our analysis. We note that 
the reconnection problem is fundamentally a boundary layer 
problem, with $S^{-1}$ being the small parameter. This allows 
us to simplify our MHD equations by performing a {\it rescaling 
procedure}\cite{Thesis} inside the reconnection layer, to make 
rescaled resistivity equal to unity. This can be done in a natural 
way if one rescales the distances and the fields in the $y$~direction 
to the corresponding {\it global} values (i.e., the length of the layer~$L$, 
the outside magnetic field just above the center of the layer $B_{y,0}(0)$, 
and the corresponding Alfv{\'e}n speed $V_A$), while rescaling the distances 
in the $x$~direction and the $x$~components of the velocity and magnetic 
field to the corresponding {\it local} values:
$$ {y\over L} \rightarrow y,\qquad \qquad 
{x\over{\delta_0}} \rightarrow x,                                   $$
$$ {v_y\over V_A} \rightarrow v_y, \qquad \qquad 
{v_x\over{V_A \delta_0/L}} \rightarrow v_x,                         $$
$$ {B_y\over{B_{y,0}(0)}} \rightarrow B_y, \qquad \qquad 
{B_x\over{B_{y,0}(0)\delta_0/L}} \rightarrow B_x,        \eqno{(4)} $$
$$ {p\over{B_{y,0}^2(0)/ 4\pi}} \rightarrow p, \qquad \qquad
{E\over{B_{y,0}(0) V_A \delta_0/L}}\rightarrow E,                   $$
where $\delta_0 \equiv L S^{-1/2}$ is the Sweet--Parker thickness of 
the current layer. Thus, one can see that the small scale~$\delta_0$ 
emerges naturally as the thickness of the resistive boundary layer. 

The viscosity $\nu$ is now rescaled as $\nu/\eta \rightarrow \nu$.
We assume that it is at least as small as the resistivity [which 
means $\nu=O(\eta)$ or less], and most of the time (see Ref.~\cite
{Uzdensky-1998}) we will be interested in the case of 
vanishing viscosity $\nu \rightarrow 0$ (which means 
$\nu \ll \eta$).

Now, all of the rescaled dimensionless quantities ($v_x$, $v_y$, $B_x$, 
$B_y$) are generally of order~1. Using the small parameter $\delta_0 /L 
= S^{-1/2} \ll 1$, one can simplify the equations by writing them down 
in the leading nontrivial order in $\delta_0 / L$. This way one neglects 
all unimportant corrections, keeping only the essential terms.

First, the incompressibility condition is written in rescaled 
quantities in exactly the same way as in the unrescaled quantities:
$$ {{\partial v_x} \over{\partial x}} + 
{{\partial v_y} \over{\partial y}} = 0.                       \eqno{(5)} $$

The $z$~component of the steady-state Ohm's law can be written as 
$$ E = {{\partial B_y}\over{\partial x}} -v_x B_y + v_y B_x,  \eqno{(6)} $$
where the first term on the RHS is the resistive term.

Next, consider the equation of motion. Since all the velocities in 
the $x$~direction are small compared with the Alfv{\'e}n speed, the 
inertial terms in the $x$~component of the equation of motion~(3) 
are small, and this equation just gives one the pressure balance 
across the current sheet:
$$ {\partial\over{\partial x}}\left(p+{B_y^2\over 2}\right)=0, \eqno{(7)} $$
which allows one to determine the pressure $p$ in terms of $B_y(x,y)$
once the pressure and the magnetic field $B_{y,0}(y)$ outside the 
reconnection layer are known. It is customary to set the pressure 
outside the layer to zero, so that
$$ p(x,y) = {{B_{y,0}^2(y)}\over 2} - {{B_y^2(x,y)}\over 2}.   \eqno{(8)} $$
(Of course, one can add a large constant to the pressure
to enforce incompressibility.  Since this constant cancels
out of all our equations, we ignore it.)

Finally, one has the $y$~component of the equation of motion, with 
acceleration provided both by the pressure gradient force,
and by magnetic
forces, and with the viscous force:
$$ {\bf v} \cdot \nabla v_y = - {{\partial p}\over{\partial y}} +
B_x {{\partial B_y}\over{\partial x}} + 
\nu {{\partial^2 v_y}\over{\partial x^2}}.                     \eqno{(9)} $$

It is interesting to note that almost everywhere in the layer 
the magnetic force $j_z B_x = B_x \partial_x B_y$ is actually
just the $y$~component of the magnetic pressure gradient force
$ \nabla_{\perp} B^2/8 \pi = \nabla B^2/ 8 \pi -
{\bf  b b} \cdot \nabla B^2/8 \pi $ where $ {\bf  b} $ is the
unit vector along $ {\bf  B } $.  The magnetic tension
force $ B^2 {\bf  b \cdot \nabla b}/ 4 \pi $ is higher order everywhere
except 
very close to the midplane (in a thin region where the unrescaled 
fields satisfy $B_y < B_x$).

We believe that this rescaling procedure captures all the important dynamical 
features of the reconnection process.

\section{Numerical Simulations}
\label{sec-simulations}

In order to find the steady-state solution for the system,
we designed a time dependent 
resistive MHD code for the main reconnection 
layer. This main code was supplemented by another code 
describing the separatrix, which is needed to provide 
the nearly correct downstream boundary conditions for the main code.

In Section~\ref{subsec-scheme} we describe the main code for 
the reconnection layer together with the boundary conditions. 
In Section~\ref{subsec-initial} we discuss the different choices 
for the initial conditions used in the simulation. In Section~\ref
{subsec-separatrix} we present the model for the separatrix region 
that we have used in order to get the downstream boundary conditions 
for the main layer. In Section~\ref{subsec-results} we report the 
results of our numerical simulations. More details can be found in 
Ref.~\cite{Thesis}.

\subsection{Numerical Scheme and Boundary Conditions}
\label{subsec-scheme}

In order to approach the steady-state solution described by the system 
of rescaled equations~(5), (6), (8), and (9), we followed the true time 
evolution of the system, starting with some initial conditions that will 
be described in Section~\ref{subsec-initial}.

The time evolution is described by the following two dynamic 
equations for the two dynamic variables $\Psi$ and $v_y$:
$$ {{\partial\Psi(x,y)}\over{\partial t}}
= -\nabla \cdot ({\bf v} \Psi) + 
{{\partial^2 \Psi}\over{\partial x^2}} + 
\left( \eta_y {{\partial^2 \Psi}\over{\partial y^2}} \right),  \eqno{(10)} $$
and
$$ {{\partial v_y}\over{\partial t}}
 = -\nabla \cdot ({\bf v} v_y) - 
{d\over{d y}} \left[{{B_{y,0}^2(y)}\over 2}\right] + 
\nabla \cdot ({\bf B} B_y) + 
\nu {{\partial^2 v_y}\over{\partial x^2}} + 
\left( \nu_y {{\partial^2 v_y}\over{\partial y^2}} \right).     \eqno{(11)} $$
These equations are written (using $\nabla \cdot {\bf v}=0$) 
in conservative form (i.e., in the form of conservation laws), 
which is preferable for numerical computations. Small artificial 
resistivity $\eta_y$ and viscosity $\nu_y$, acting in the $y$~direction,
are added to provide numerical stability. Because they are small, 
these terms do not change the solution noticeably, as was verified 
in the runs. The natural unit of time in our
simulations is the global Alfv{\'e}n time $\tau_A = L/V_A$.

Once $v_y$ and $\Psi$ are known everywhere at a new
time step, one can find all other variables. Namely, 
$B_x$ and $B_y$ are given by the derivatives of $\Psi$:
$$ B_x = -{{\partial \Psi}\over{\partial y}}, \qquad
B_y = {{\partial \Psi}\over{\partial x}},                    \eqno{(12)} $$
and $v_x$ is obtained from the incompressibility condition
by integrating $\partial v_y /\partial y$ in the $x$~direction
starting from the midplane, where $v_x=0$ because of the
symmetry:
$$ v_x (x,y) = -\int_0^x{{\partial v_y}\over{\partial y}}dx. \eqno{(13)} $$

We used the finite-difference method with centered derivatives 
(providing second order accuracy) in both the $x$ and the 
$y$~directions. 

The time derivatives were one-sided. Our scheme was explicit in the 
$y$~direction, but in the $x$~direction the resistive term in Ohm's 
law ($\partial^2 \Psi / \partial x^2$) was treated implicitly, while 
all other terms were treated explicitly. This enabled us to speed up 
the computations.

\begin{figure} [t]
\centerline {\psfig{file=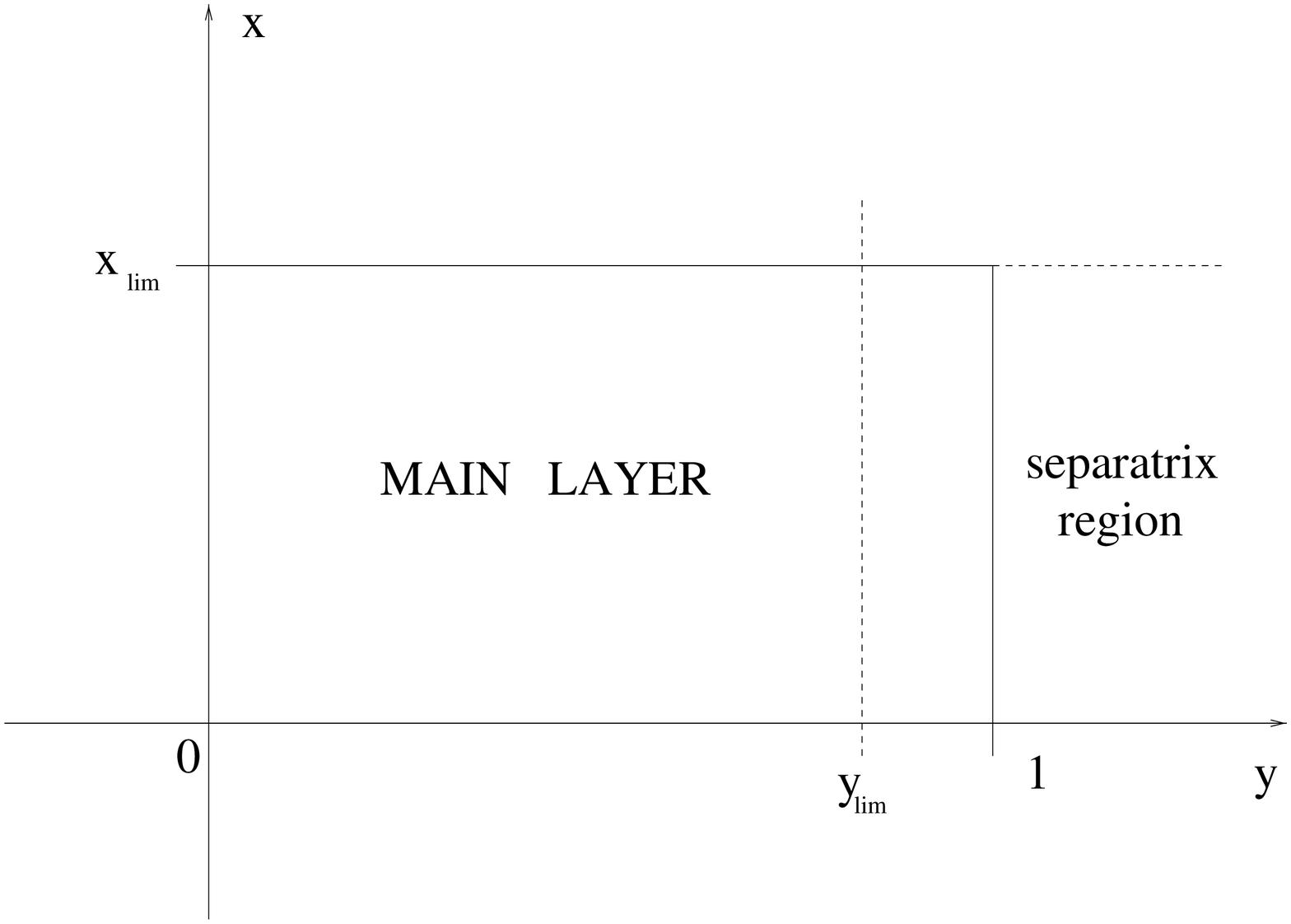,height=3 in,width=5 in}}
\caption[Computational box used in the numerical simulation]
{Computational box used in the numerical simulation.}
\label{fig-comp-box}
\end{figure}

We conducted the simulations on a rectangular grid ($i_m \times 
j_m$). Because of the symmetry, we considered only one quadrant 
(see Fig.~\ref{fig-comp-box}).

Now let us discuss the boundary conditions. There are four boundaries 
in the system: the upstream (or upper) boundary $x=x_{\rm lim}$, 
$0<y<y_{\rm lim}$, the downstream (or right) boundary $y=y_{\rm lim}$, 
$0<x<x_{\rm lim}$, and the two boundaries formed by the axes of symmetry:
the lower boundary $x = 0$, $0 < y < y_{\rm lim}$, and the left boundary 
$y=0$, $0<x<x_{\rm lim}$. The flow enters through the upper boundary and 
leaves through the right boundary, so that there is no flow of plasma 
through the left and the lower boundaries. While the boundary conditions 
at the left and the lower boundaries come from simple symmetry conditions, 
the boundary conditions on the upper boundary and especially on the right 
boundary are more complicated, as we shall discuss below.

On the upper boundary, the boundary conditions come from matching 
with the ideal MHD solution in region~I above the reconnection layer. 
In rescaled quantities, this matching should be done at $x\rightarrow 
\infty$. But since in a numerical simulation it is not possible to 
place a boundary of the computation box at infinity, we place it at 
some sufficiently large $x_{\rm lim} \gg 1$. The typical values of 
$x_{\rm lim}$ in our simulations were $x_{\rm lim}\simeq 4-8$.

From the ideal solution in region~I we know that, as $x \rightarrow \infty$, 
$v_y \rightarrow 0$ (meaning $v_y \ll V_A$) and $B_y \rightarrow B_{y,0}(y)$, 
which is prescribed. It turns out that, since the upper boundary is placed 
not at infinity but at some finite (although large) $x_{\rm lim}$, it is 
better, for numerical reasons, to choose ${\partial v_y} / {\partial x}
(x_{\rm lim})=0$ instead of $v_y(x_{\rm lim}) = 0$. 

As for the magnetic field, we just set 
${\partial\Psi}/{\partial x}(x_{\rm lim})=B_{y,0}(y)$. 
The function $B_{y,0}(y)$ depends on the external boundary 
and the thermodynamics of the global solution as well as 
the amount of reconnected flux.  Rather than make use of 
a specific global solution  we choose a generic one since 
the physics in the reconnection layer should be the same.
However, near $y=L$ (the end of the reconnection layer 
which terminates in a cusp geometry we must employ the 
square-root behavior derived in Ref.~\cite{Uzdensky-1997},  
which any global solution must have. Thus, in our numerical 
simulations we typically took
$$ B_{y,0}(y) = B_0 + (1-B_0) \sqrt{1-y^2},                  \eqno{(14)} $$
consistent with the cusp solution (see Ref.~\cite{Uzdensky-1997}). 
The value $B_0$ of the outside magnetic field $B_{y,0}$ at the 
endpoint $y=1$ typically was taken to be $B_0=0.2$ or $B_0=0.3$ 
[the magnetic field is normalized so that $B_{y,0}(0)=1$]. 

This choice of boundary conditions worked well in our simulations.
In particular, the behavior of the solution near the upper boundary
was smooth, and the solution deep inside the reconnection layer did
not depend on the exact position of the upper boundary.

At the lower boundary $x=0$, the boundary conditions come naturally 
from the requirement that both $v_y$ and $\Psi$ be symmetric with 
respect to the midplane. Thus, 
$$ {{\partial v_y}\over{\partial x}} =0 \qquad {\rm and}
\qquad B_y={{\partial \Psi}\over{\partial x}}=0 \qquad (x=0), \eqno{(15)} $$
and also $v_x(0,y)=0$.

Here, however, we would like to make one remark. The boundary condition 
for ${\partial v_y/}{\partial x}$ at $x=0$ is needed only when one includes 
the viscous term $\nu\partial^2 v_y/\partial x^2$ in the equation of motion. 
If one does not keep this term, then this equation contains only first 
derivatives of $v_y$ in the $x$~direction, so one needs only one condition, 
which can be set at the upper boundary. In our simulations, however, we 
always include viscosity (usually small, but not zero), both for numerical 
reasons and in order to resolve the behavior near the midplane (see 
Ref.~\cite{Uzdensky-1998}).

The boundary conditions at the left boundary are similar to
those on the lower boundary. They follow from symmetry and are 
rather straightforward:
$$ v_y (x,0) = 0, \qquad {\rm and} \qquad B_x(x,0)=0.        \eqno{(16)} $$

Now consider the right (or downstream) boundary.  We chose it at some 
point $y = y_{\rm lim}$ close to the endpoint (typically $y_{\rm lim} 
= 0.9-1.0$). Then, one needs to specify the downstream boundary 
conditions on this boundary. This boundary is, in fact, the interface 
between the main layer and the endpoint region and the separatrix. The 
boundary conditions should describe the effect of the separatrix reacting 
back on the main layer, in particular the back pressure. We are not aware 
of any previous numerical or theoretical studies in which the role of the 
back pressure has been adequately investigated. The problem of how to set 
the boundary conditions on this boundary is rather nontrivial and its 
discussion is postponed until Section~\ref{subsec-separatrix}.

To summarize, the advantages of our approach to numerical 
simulation of the reconnection layer are the following:

1) First, the use of rescaled equations takes us directly
into the realm of $S \rightarrow \infty$.

2) Second, this is an essentially 2D (rather than 1D) code that
uses a realistic variation of the outside magnetic field along 
the layer. The position of the endpoint is clearly defined in 
terms of the function $B_{y,0}(y)$. We do not assume $B_{y,0}(y)
={\rm const}$ as many people do.

3) We obtain the steady-state solution by following the true time 
evolution, and the rescaled equations are such that we do not give 
the boundary conditions for the incoming flow velocity at the upstream 
boundary. This is because the physical $v_x$ is small and its evolution 
is not determined from a dynamic equation of motion. Instead, the equation 
of motion in the $x$~direction simply degenerates into the vertical pressure 
balance, and $v_x$ is determined from the incompressibility condition. Thus, 
we do not specify $v_x(x_{\rm lim},y)$ as a boundary condition, which means 
that \emph{we do not prescribe the reconnection rate}! The system itself 
determines what the reconnection rate should be! This is really a very 
important point.

The fact that we rescaled the $x$~coordinate using the Sweet--Parker scaling 
does not actually mean that we prescribe the Sweet--Parker reconnection rate. 
If the system wants to go at a faster rate, then it would try to develop 
some new characteristic structures, extending beyond $x_{\rm lim}$, and 
we should be able to see it.

\subsection{Initial Conditions}
\label{subsec-initial}

We have performed several runs with different initial conditions consistent 
with our boundary conditions.

\begin{figure} [tbp]
\centerline
{\psfig{file=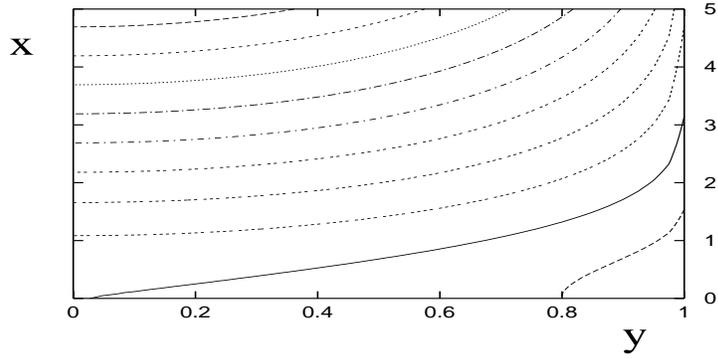,height=3in,width=6in}}
\caption[Contour plot of magnetic field for Sweet--Parker-like initial 
conditions]
{Contour plot of the magnetic field for Sweet--Parker-like initial conditions.}
\label{fig-local-init-Psi-cont-SP1}
\end{figure}

\begin{figure} [tbp]
\centerline
{\psfig{file=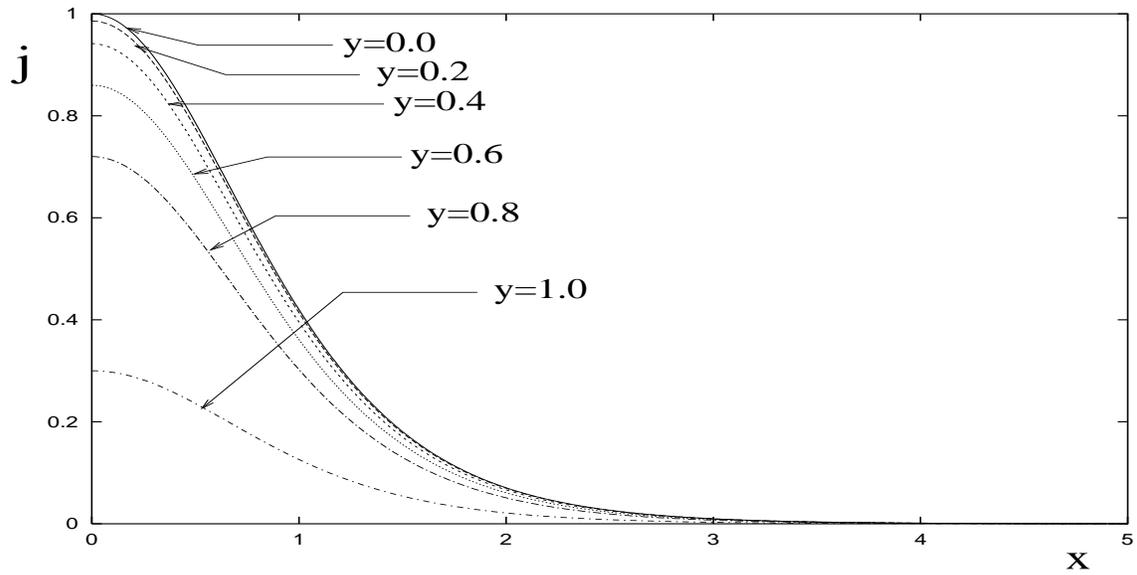,height=3in,width=6in}}
\caption[Plots of $j(x)$ for different values of $y$ at $t=0$
for Sweet--Parker-like initial conditions]
{Plots of $j(x)$ for different values of $y$ at $t=0$
for Sweet--Parker-like initial conditions.}
\label{fig-local-init-j-x-SP1}
\end{figure}

In some cases we started with a configuration qualitatively resembling the 
Sweet--Parker reconnection layer (see Figs.~\ref{fig-local-init-Psi-cont-SP1}
and \ref{fig-local-init-j-x-SP1}). These initial conditions can be written 
in the following analytical form:

$$ B_y(x,y) = B_{y,0}(y) \tanh(x),                         \eqno{(17)} $$
$$ \Psi(x,y) = \Psi(0,y) + B_{y,0}(y) \log{\cosh{x}},      \eqno{(18)} $$
$$ B_x(x,y) = - {{\partial \Psi}\over{\partial y}},        \eqno{(19)} $$
where we took the outside magnetic field $B_{y,0}(y)$ in the form~(14),
and the variation of the magnetic flux on the midplane as
$$ \Psi(0,y) = - {\pi\over 4} y^2.                         \eqno{(20)} $$

The initial velocity was taken in the form
$$ v_x(x,y) = - {E_0\over{B_{y,0}(y)}} \tanh(x),           \eqno{(21)} $$
and 
$$ v_y(x,y)=-\int_0^y{{\partial v_x}\over{\partial x}}dy,  \eqno{(22)} $$
where the initial outside electric field $E_0$ varied but typically 
was of order one. We call the initial conditions described by the set 
of Eqs.~(4.17)--(4.22) the Sweet--Parker-like initial conditions.

In other runs we wanted to see whether the system would want to go 
with the faster reconnection rate and whether it would develop the
Petschek-like structures. 

(It should be observed that Petschek, in his original paper, did 
not specify a unique reconnection rate. Instead his reconnection 
rate depended on the length of the dissipation region $y_*$, and 
his reconnection velocity was $v_{\rm rec}=(\sqrt{ L/y_*}) V_A /
\sqrt{S}$. For example, the choice $y_*=L$ gives the Sweet-Parker 
rate. The usually quoted Petschek reconnection velocity, $v_{\rm 
Petschek}\sim V_A/ \log{S}$, was obtained by him when he took the 
minimum  possible value for $y_*$ that would not seriously perturb 
the global solution. Furthermore, it is possible to include Petschek's 
shocks inside the boundary layer as indeed he did.)

\begin{figure} [tbp]
\centerline
{\psfig{file=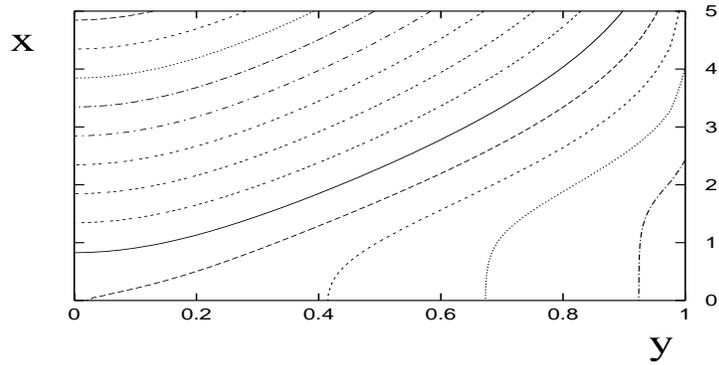,height=3in,width=6in}}
\caption[Contour plot of magnetic field for Petschek-like initial conditions]
{Contour plot of magnetic field at $t=0$ for the Petschek-like initial 
conditions.}
\label{fig-local-init-Psi-cont-petschek}
\end{figure}

\begin{figure} [tbp]
\centerline
{\psfig{file=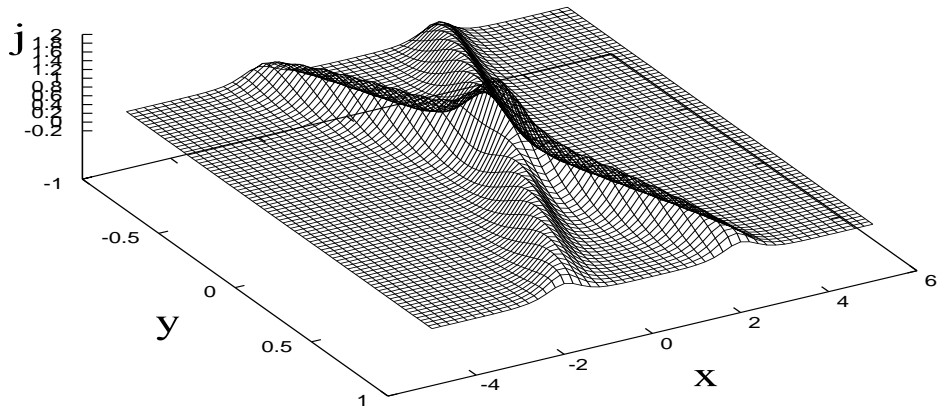,height=3in,width=6in}}
\caption[The current density $j(x,y)$ at $t=0$ for Petschek-like initial 
conditions]
{The current density $j(x,y)$ at $t=0$ for Petschek-like initial 
conditions (all four quadrants are shown for clarity).}
\label{fig-local-init-j-3d-petschek}
\end{figure}

\begin{figure} [tbp]
\centerline
{\psfig{file=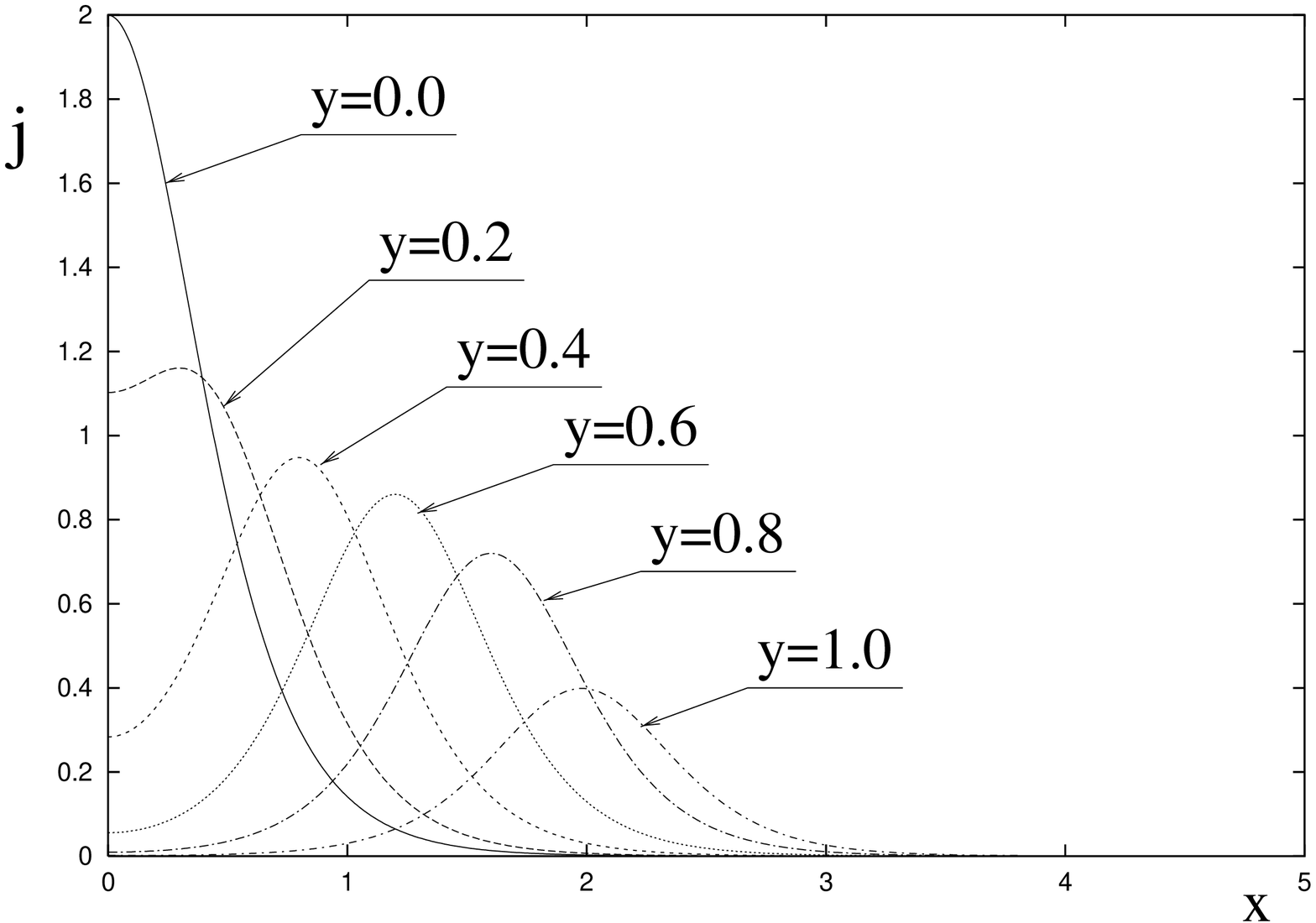,height=3in,width=6in}}
\caption[Plots of $j(x)$ for different values of $y$ at $t=0$
for Petschek-like initial conditions]
{Plots of $j(x)$ for different values of $y$ at $t=0$
for the Petschek-like initial conditions.}
\label{fig-local-init-j-x-petschek}
\end{figure}

We conjectured that if the system wants to go at a faster reconnection 
rate than that of Sweet--Parker, then it would at least be able to go 
at a rate twice as fast({\it i.e.}, $y_* = L/4$). Therefore, we carried 
out several runs where at $t=0$ we set up a Petschek-like structure (see 
Figs.~\ref{fig-local-init-Psi-cont-petschek},
\ref{fig-local-init-j-3d-petschek}, 
and \ref{fig-local-init-j-x-petschek}) 
described by the following expressions: 
$$ \Psi(0,y)=-{1\over M} \log{\cosh(z_0)},                   \eqno{(23)} $$
$$ B_y(x,y) = B_{y,0}(y) \left[{{\tanh(z_1)+\tanh(z_2)} 
\over 2} \right],                                            \eqno{(24)} $$
$$ \Psi(x,y)=\Psi(0,y)+{{B_{y,0}(y)}\over{2M}}\
[\log{\cosh(z_1)} + \log{\cosh(z_2)} - 2\log{\cosh(z_0)}],   \eqno{(25)} $$
$$ B_x(x,y) = -  {{\partial \Psi}\over{\partial y}},         \eqno{(26)} $$
$$ v_y(x,y)= {{\tanh(z_1)-\tanh(z_2)}\over 2},               \eqno{(27)} $$
$$ v_x(x,y) = - M \left[{{\tanh(z_1)+\tanh(z_2)}
\over 2} \right],                                            \eqno{(28)} $$
where $z_0=M^2 y$, $z_1=Mx+M^2 y$, and $z_2=Mx-M^2 y$, and where 
the parameter~$M$ corresponds to the initial reconnection rate in 
terms of the Sweet--Parker reconnection rate. It describes how well 
pronounced the Petschek-like structure is ($M$ would correspond to 
$\sqrt{L/y_*}$ in Petschek's notation). Typically $M$ was chosen to 
be~2 or~3. We call the initial conditions described by the set of 
Eqs.~(23)--(28) the Petschek-like initial conditions.

\subsection{The Downstream Boundary Conditions and the Model 
for the Separatrix Region}
\label{subsec-separatrix}

The downstream boundary $y=y_{\rm lim}$ is the interface between 
the main layer and the separatrix region. The boundary conditions 
at this boundary cannot be given in a simple closed form.%
\footnote
{For example, we have tried to use the so-called free-flow boundary 
conditions: $\partial^2 \Psi / \partial y^2=0$, $\partial^2 v_y / 
\partial y^2=0$. The steady-state solution exists and is reached 
within several Alfv{\'e}n times. However, the solution in the bulk 
of the main layer strongly depends on the position of both the upstream 
and downstream boundaries, which is physically not acceptable.}
Instead, they require matching with the solution in the separatrix 
region, which itself is just as complicated as the main layer. 
Therefore, we have developed a supplemental numerical procedure 
for the separatrix region. 

To make the situation more tractable, we have adopted
\emph{simplified physical model for the separatrix} --- 
namely, a model in which the resistive and the viscous 
terms are omitted, and hence the magnetic field is frozen 
into the plasma. Even though this model does not describe 
the separatrix completely accurately, it should give one 
a qualitatively correct picture of the influence of the 
separatrix region back on the main layer, and thus sufficiently 
reasonable downstream boundary conditions for the main layer. 
In particular, our model includes the effects of the back pressure 
that the separatrix exerts on the main layer.

We follow one given field line as it moves across the separatrix of 
length~$L_s$ as described in the introduction. This field line is 
described in terms of two functions: the magnetic field $B(l,t)$, 
and the parallel velocity $v(l,t) \equiv v_{\parallel}(l,t)$. Here 
$l$ is the length measured along the field line, starting from the 
boundary between the separatrix and the main layer ($l=0$ or 
$y=y_{\rm lim}$), and ending at the reflection point~A at the 
top of Fig.~\ref{fig-global} ($l=L_s$).

Now, the time evolution in this one-dimensional problem corresponds to 
the perpendicular motion of the field line through the separatrix region. 
The new time variable~$t$ represents the relative position of the field 
line in real 2D space with respect to other field lines, and there is a 
one-to-one correspondence between the time~$t$ and the $x$~position of 
the footpoint (the point where the field line intersects the boundary 
$y=y_{\rm lim}$) of this field line. This correspondence between $t$ 
and~$x$ is given by the $x$~component of the velocity of the field line, 
$v_{xs} \equiv - E/B_y(x,y_{\rm lim})<0$ ($x$ decreases as $t$ increases), 
and by the initial condition that $x=x_{\rm lim}$ at $t=0$. We stop the 
simulation of the separatrix at the moment $t$ when $x(t)$ becomes zero 
--- i.e., when the midplane is reached. A substantial advantage of this 
model is that one actually does not need to know where the field line is 
in real 2D space! The only thing one does need to know about the position 
of the field line in order to set up the downstream boundary conditions 
for the main layer is where it connects to the main layer --- i.e., what 
is the value of the coordinate $x$ at the point where the field line 
intersects the downstream boundary $y=y_{\rm lim}$ (i.e., the footpoint).

The boundary conditions on the upper boundary of the separatrix are 
$v(l)=0$ and $B(l)=B_{s1}(l)$, where $B_{s1}(l)$ is the field outside 
the separatrix in the upstream region given by the global ideal 
MHD solution, as explained in Ref.~\cite{Uzdensky-1997}. In our 
simulation we took this outside field in the form 
$$ B_{s,I}(l) = \sqrt{B_0^2+
{{C^2 l}\over L_s} \left( 1-{l\over{2L_s}} \right) }.         \eqno{(29)} $$ 

Now these conditions become the \emph{initial} conditions
at $t=0$ for our 1D (plus time) problem concerning this 
one field line.

As in the main layer, we use here an explicit (in $l$) code with 
centered differences for convective derivatives in the $l$~direction. 
The boundary conditions at the right boundary $l=L_s$ (the reflection 
point~A on Fig.~\ref{fig-global}) come from the condition of symmetry 
with respect to this point ($l=L_s$):
$$v(L_s,t)=0,\qquad {{\partial B}\over{\partial l}}(L_s,t)=0. \eqno{(30)} $$

The other pair of boundary conditions is given at the left (incoming)
boundary $l=0$ and is provided by the main layer in terms of the values 
of $v_y(x,y_{\rm lim})$ and $B_y(x,y_{lim})$. Assuming that the steady 
state in the main layer has already been achieved, these boundary conditions 
are constant in time in the laboratory frame, but in the frame moving along
together with the field line, they are now time-dependent.

Since at the boundary $y=y_{\rm lim}$ the layer is still essentially straight, 
and the field lines (except those few very close to the midplane, which are 
ignored) are also almost straight, then the absolute value of the magnetic 
field $B$ is almost exactly equal to~$B_y$ and the parallel velocity $v$ 
is almost exactly equal to~$v_y$. Therefore, the matching conditions at 
this boundary are $B_y=B$, $v_y=v$. 

As for the downstream (or lower) boundary, one really does not need 
to set any conditions there because the $x$~coordinate in our model 
corresponds to time and the new equations have only first-order time 
derivatives.

Now let us derive the differential equations for this model of the 
separatrix. The main idea in this derivation is that the convective 
term ${\bf v}_{\perp} \cdot \nabla_{\perp}$ is replaced by the time 
derivative $\partial / \partial t$. The natural unit of time in this 
model is the Alfv{\'e}n time, the natural unit of distance in the 
$l$~direction is the global scale~$L$, and that of the parallel 
velocity is the Alfv{\'e}n velocity~$V_A$. 

First of all, one can apply this model only to the part of the separatrix
where the field lines are not very strongly curved (i.e., where the radius 
of curvature of the field lines is of order~$L$), the magnetic field itself 
is sufficiently strong (of order $B_0$ or at least $B \gg B_1 = E/V_A$), and 
the perpendicular velocity $v_{\perp}$ (which in ideal MHD is equal to $E/B$)
is small compared with the parallel velocity $v_{\parallel}=O(V_A)$. Thus, 
the model is bound to fail in the very small region near the midplane where 
the real physical $B_y$ becomes comparable with $B_x$.%
\footnote 
{This presumably occurs at some infinitesimally short distance 
$\delta_1$ from the midplane, where $\delta_1/\delta_0\sim\delta_0 
/L \ll 1$. Since one considers the limit $S \rightarrow\infty$, and 
hence $\delta_0/L\rightarrow 0$, this region shrinks to zero, and one 
should not be concerned too much about it in our simulation.
} 

In the system of reference moving together with the field line
in the direction perpendicular to ${\bf B}$, the (parallel component 
of) the magnetic induction equation becomes
$$ \dot{B} = {{\partial B}\over{\partial t}} =
B{\partial\over{\partial l}}v-v{\partial\over{\partial l}}B,  \eqno{(31)} $$
and the parallel equation of motion becomes
$$ {{\partial v}\over{\partial t}} = -v{{\partial v}\over{\partial l}}-
{\partial\over{\partial l}} \left[{B_{s1}^2(l)\over 2}-{B^2(l,t)\over 2}
\right].                                                      \eqno{(32)} $$
Eq.~(32) is equivalent to Eqs.~(7) and~(11). Physically, the force 
parallel to ${\bf B}$ is the pressure gradient force which can be 
expressed in terms of $B$ by Eq.~(7). It is interesting to note that, 
while the fluid in the initial problem was incompressible, in this 1D 
problem the motion is effectively not incompressible (${\partial v/
\partial l} \neq 0$), and the thickness of the flux tube ($\sim 1/B$) 
plays the role of density.

Now let us see how the back pressure from the separatrix acts on
the main layer. 

As the field line moves, the incoming velocity $v(l=0,t)$ increases 
from almost zero at $t=0$ to about the Alfv{\'e}n velocity, and the 
incoming magnetic field $B(l=0, t)$ drops from $B_0$ to zero at the 
midplane. Therefore, there is a point $x=x_c$ somewhere in the middle 
of the left boundary of the separatrix where the incoming flow becomes 
locally super-Alfv{\'e}nic ($v>B$). The propagation of information with 
respect to the fluid in our model occurs at the local Alfv{\'e}n speed
by the means of two characteristics $dl/dt=v \pm B$. For $x>x_c$ ($v<B$), 
one characteristic goes from the left boundary $l=0$ ($y=y_{\rm lim}$) 
towards the right boundary, while the other characteristic goes from 
the upper boundary $x=x_{\rm lim}$ towards the left boundary, carrying 
information about the pressure and the flow in the separatrix to the 
main layer. This means that, for $x > x_c$, the layer ``feels'' the 
effect of the separatrix in the form of the back pressure coming from 
the previously undisturbed fluid in the separatrix region. This back 
pressure is found to have some stabilizing effect, making the solution 
deep inside the main layer independent of $x_{\rm lim}$ and almost 
independent of the position $y_{\rm lim}$ of the downstream boundary 
(where the separatrix and the main layer solutions are matched).

After the field line crosses the point $x=x_c$, the flow becomes 
super-Alfv{\'e}nic and the characteristic $dl/dt=v-B>0$ is deviated
away from the left boundary back into the separatrix region. Now 
both characteristics come out of the left boundary and there is no 
propagation of information from the separatrix region into the main 
layer. This also means that when the field line reaches the vicinity 
of the midplane $x=0$ and the description of the separatrix region as 
a region of almost straight field lines fails, it does not matter much, 
because the flow at this point is very strongly super-Alfv{\'e}nic and 
the main layer does not feel what happens downstream. We would like to 
remark that in our simulations we have not observed any shock formation
in the separatrix.

Of course, this picture of the flow of information is valid only
in this ideal MHD model. In the real situation with resistivity, 
there is always propagation of information upstream due to the 
resistive diffusion.

To summarize, the model of the separatrix region presented 
here has two main drawbacks. First, it describes an ideal 
MHD separatrix. Second, it {\it assumes} that the boundary 
conditions on the incoming boundary $l=0$ ($y=y_{\rm lim}$) 
are stationary (i.e., that the main layer has already reached 
the steady state). (This second objection can be removed if 
the main layer is evolved with right hand boundary conditions 
changing in step with the separatrix solution and {\it vice 
versa}, but this would involve a longer simulation.) In addition, 
the model is valid only when the field lines are essentially straight, 
as discussed above (the region where this assumption breaks down is 
discussed in Ref.~\cite{Thesis}). Despite all that, however, we feel 
that this model provides a \emph{qualitatively correct picture of the 
dynamics in the separatrix region in the steady state}, and thus gives 
us sufficiently reasonable downstream boundary conditions for the main 
layer.

\subsection{Results of the Simulations}
\label{subsec-results}

Now let us present and discuss the results of our numerical simulations.

We found that, after a transient period of a few Alfv{\'e}n times, 
\emph{the system reaches a steady state that is independent of the 
initial configuration}.

In particular, when we start with a Petschek-like initial configuration 
(described in Section~\ref{subsec-initial}), the high velocity 
flow rapidly sweeps away the transverse magnetic field~$B_x$ (see 
Fig.~\ref{fig-Bx(0,y)-t-petschek}). This is important, because, for 
a Petschek-like configuration to exist, the transverse component of 
the magnetic field on the midplane, $B_x(0,y)$, must be large enough 
to be able to sustain the Petschek shocks in the field reversal region.
For this to happen, has to rise rapidly with~$y$ inside a very short diffusion 
region, $B_x(0,y)$ $y < y_* \ll L$ (in the case $M = 2$, presented in 
Fig.~\ref{fig-Bx(0,y)-t-petschek}, $y_* = L/4$), to reach a certain large 
value ($B_x=2$ for~$M=2$) for $y_* \ll y < L$. While the transverse magnetic 
flux is being swept away by the plasma flow, it is being regenerated by the 
merging of the $B_y$~field, but only at a certain rate and only on a global 
scale in the $y$-direction, related to the nonuniformity of the outside 
magnetic field~$B_{y,0}(y)$, as discussed in Refs.~\cite{Kulsrud-1998} 
and~\cite{Thesis}. As a result, the initial Petschek-like structure is 
destroyed, and the inflow of the magnetic flux through the upper boundary 
drops in a fraction of one Alfv{\'e}n time. Then, after a transient period, 
the system reaches a steady state consistent with the Sweet--Parker model.

\begin{figure} [t]
\centerline 
{\psfig{file=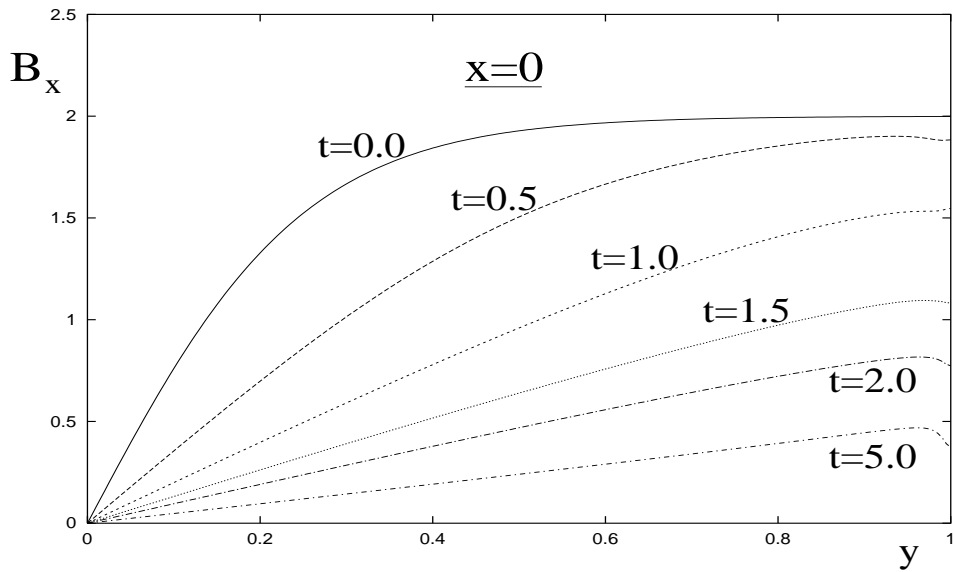,height=3.in,width=5in}}
\caption[The time evolution of the variation of the transverse magnetic 
field $B_x(0,y)$ along the midplane $x=0$ for the Petschek-like initial 
conditions]
{The time evolution of the variation of the transverse magnetic field 
$B_x(0,y)$ along the midplane $x=0$ for the Petschek-like initial conditions.}
\label{fig-Bx(0,y)-t-petschek}
\end{figure}

\begin{figure} [tbp]
\centerline {\psfig{file=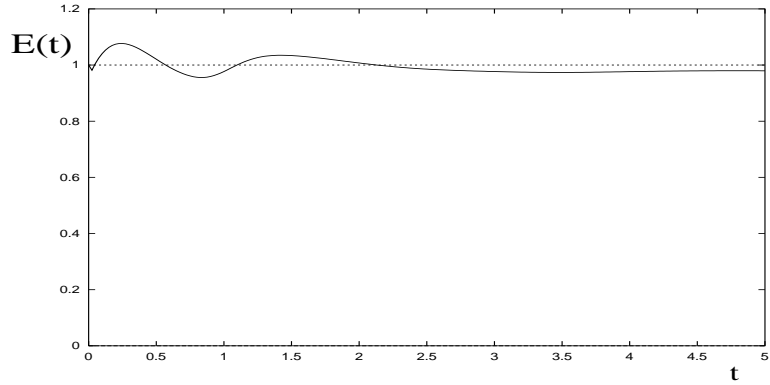,height=2.in,width=4in}}
\caption[Electric field at $x=x_{\rm lim}$, $y=y_{\rm lim}$ as a function 
of time for the Sweet--Parker-like initial conditions with $E_0=1.0$]
{The time evolution of the electric field $E$ at the point $x=x_{\rm lim}$, 
$y=y_{\rm lim}$ for the Sweet--Parker-like initial conditions with $E_0=1.0$.}
\label{fig-local-E(t)-SP1}
\end{figure}

\begin{figure} [tbp]
\centerline {\psfig{file=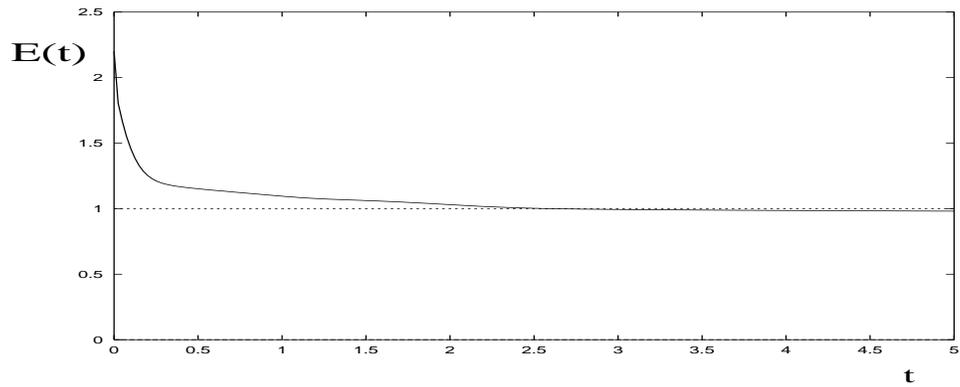,height=2.in,width=5in}}
\caption[Electric field at $x=x_{\rm lim}$, $y=y_{\rm lim}$ as a function of 
time for the Petschek-like initial conditions]
{The time evolution of the electric field $E$ at the point $x=x_{\rm lim}$, 
$y=y_{\rm lim}$ for the Petschek-like initial conditions with $M=2$ 
(corresponding to $E_0=2.0$).}
\label{fig-local-E(t)-petschek}
\end{figure}

In general, the transient period typically lasts a few Alfv{\'e}n 
times, during which the incoming electric field can oscillate around 
its final steady-state value. These oscillations have a period of order 
$\tau_A$, and a decay time also of the same order. After several $\tau_A$, 
the electric field becomes constant and uniform throughout the computational 
domain and the system approaches steady state. We terminate our simulations 
typically after 5 or 10 Alfv{\'e}n times. The time evolution of the incoming 
electric field [i.e., $E(x=x_{\rm lim},0)(t)$] for different choices of the 
initial conditions is represented in Figs.~\ref{fig-local-E(t)-SP1}  
and~\ref{fig-local-E(t)-petschek}.

We believe that the fact that we rescaled $x$ using the Sweet--Parker 
scaling does not mean that we prescribe the Sweet--Parker reconnection 
rate. Indeed, if the reconnecting system wanted to evolve towards 
Petschek's fast reconnection, it would then try to develop some new 
characteristic structures, e.g., Petschek-like shocks, which we would 
be able to see. Note that, if Petschek is correct, then there should 
be a range of reconnection rates including those equal to any finite 
factor greater than one times the Sweet--Parker rate~$E_{\rm SP}$. 
However, in our simulations we have demonstrated that there is only 
one stable solution and that it corresponds to~$E=E_{\rm SP}$. In this 
sense we have demonstrated that Petschek must be wrong since reconnection 
can not even go a factor of two faster than Sweet--Parker, let alone almost 
the entire factor of~$\sqrt{S}$. There seems no alternative to the conclusion 
that fast reconnection is impossible. 

It is interesting that in Petschek's original paper~\cite{Petschek-1964}
the length of the central diffusion region $y_*$ is an 
undetermined parameter, and the reconnection velocity 
$v_{\rm rec}$ depends on this parameter as $V_A (L/y_*)^2 /
\sqrt{S} $. If $y_*$ is taken as small as possible then 
Petschek finds that $v_{\rm rec} \sim V_A/ \log(S)$.
However, $y_*$ should be determined instead by balancing 
the generation of the transverse field $B_x$ against its 
loss by the Alfv{\'e}nic flow (it should be remarked that
Petschek did not discuss the origin of this transverse
field in his paper). As we discussed above, this balance 
yields $y_* \approx L$, with the resulting unique rate 
equal to that of Sweet--Parker. This results are borne 
out by our time-dependent numerical simulations.

The final steady-state configuration represents the \emph{Sweet--Parker 
reconnection layer}. This means that all the plasma parameters are of 
order one in the rescaled coordinates, and change on a scale of order 
$\delta_0$ in the $x$~direction and on a global scale~$L$ in the 
$y$~direction. For our choice of boundary conditions [i.e., of the 
outside magnetic field $B_{y,0}(y)$], the reconnection rate in the 
steady state was $E=1.0\, E_{\rm SP}$, where $E_{\rm SP} \equiv 
\eta^{1/2} V_A B_{y,0}(0)$ is the typical Sweet--Parker reconnection 
rate.

The use of the time-dependent equations allows us not only to find 
the steady-state solution, but also to draw some conclusions about 
its stability. The Sweet--Parker solution was found to be stable 
and robust: it did not depend on the positions of the boundaries 
$x_{\rm lim}$, $y_{\rm lim}$ or on the small artificial resistivity 
and viscosity. Moreover, we found that it is fairly insensitive even 
with respect to the choice of the parameters describing the outside 
magnetic field, such as $B_0=B_{y,0}(1)$ (thus, $E$ varies by about 
10\% as $B_0$ changes from 0 to 0.3).

\begin{figure} [t]
\centerline
{\psfig{file=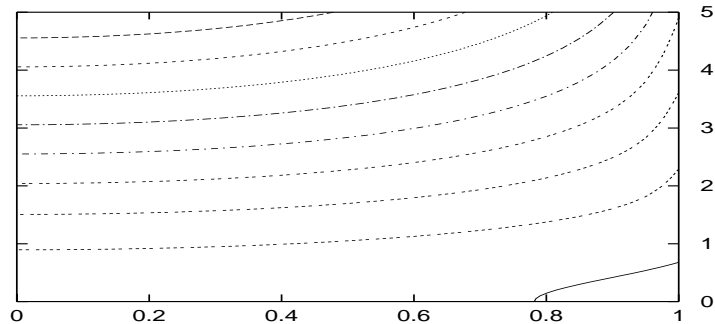,height=3in,width=6.5in}}
\caption[Contour plot of the magnetic field in the final steady state]
{Contour plot of the steady-state magnetic field in the reconnection layer.}
\label{fig-local-steadystate-Psi-cont}
\end{figure}

\begin{figure} [tbp]
\centerline
{\psfig{file=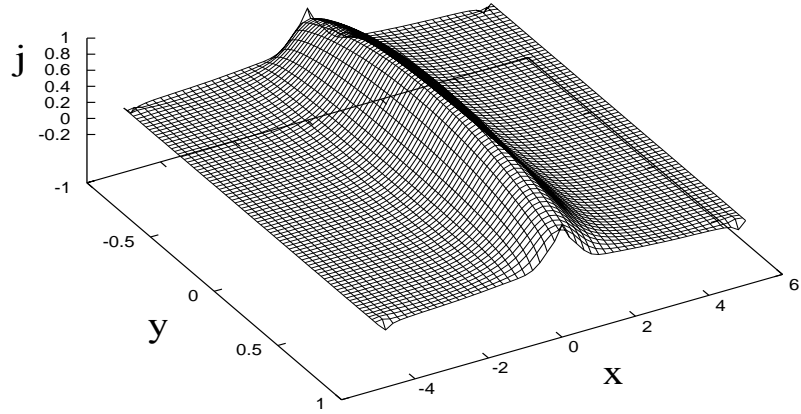,height=3in,width=5in}}
\caption[The current density $j(x,y)$ in the steady state]
{The current density $j(x,y)$ in the steady state (all four quadrants 
are shown for clarity).}
\label{fig-local-steadystate-j-3d}
\end{figure}

\begin{figure} [tbp]
\centerline
{\psfig{file=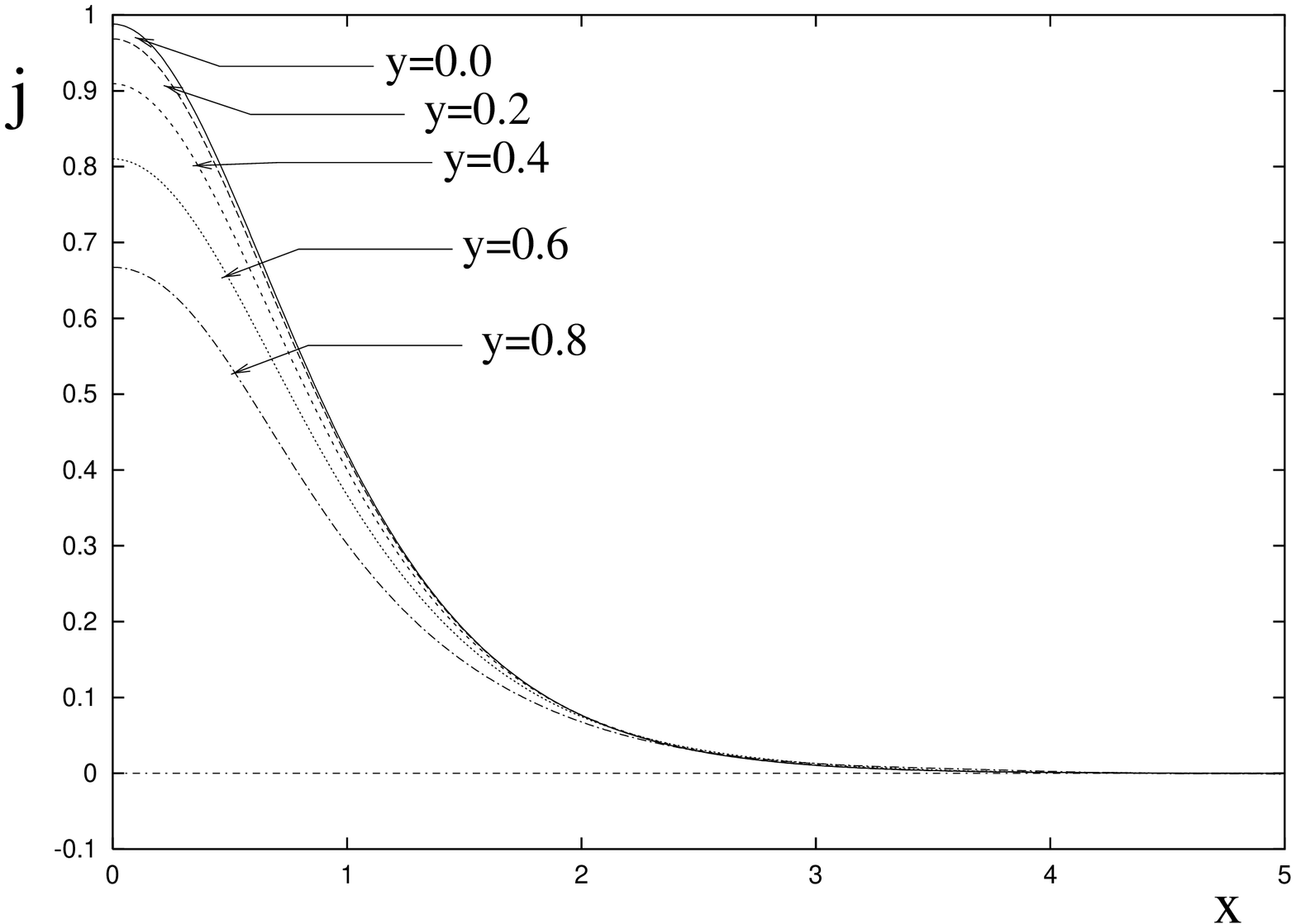,height=3in,width=5in}}
\caption[Plots of $j$ as a function of $x$ for several different
values of $y$ in the steady state]
{Plots of $j$ as a function of $x$ for several different values of 
$y$ in the steady state.}
\label{fig-local-steadystate-j(x)}
\end{figure}

\begin{figure} [tbp]
\centerline
{\psfig{file=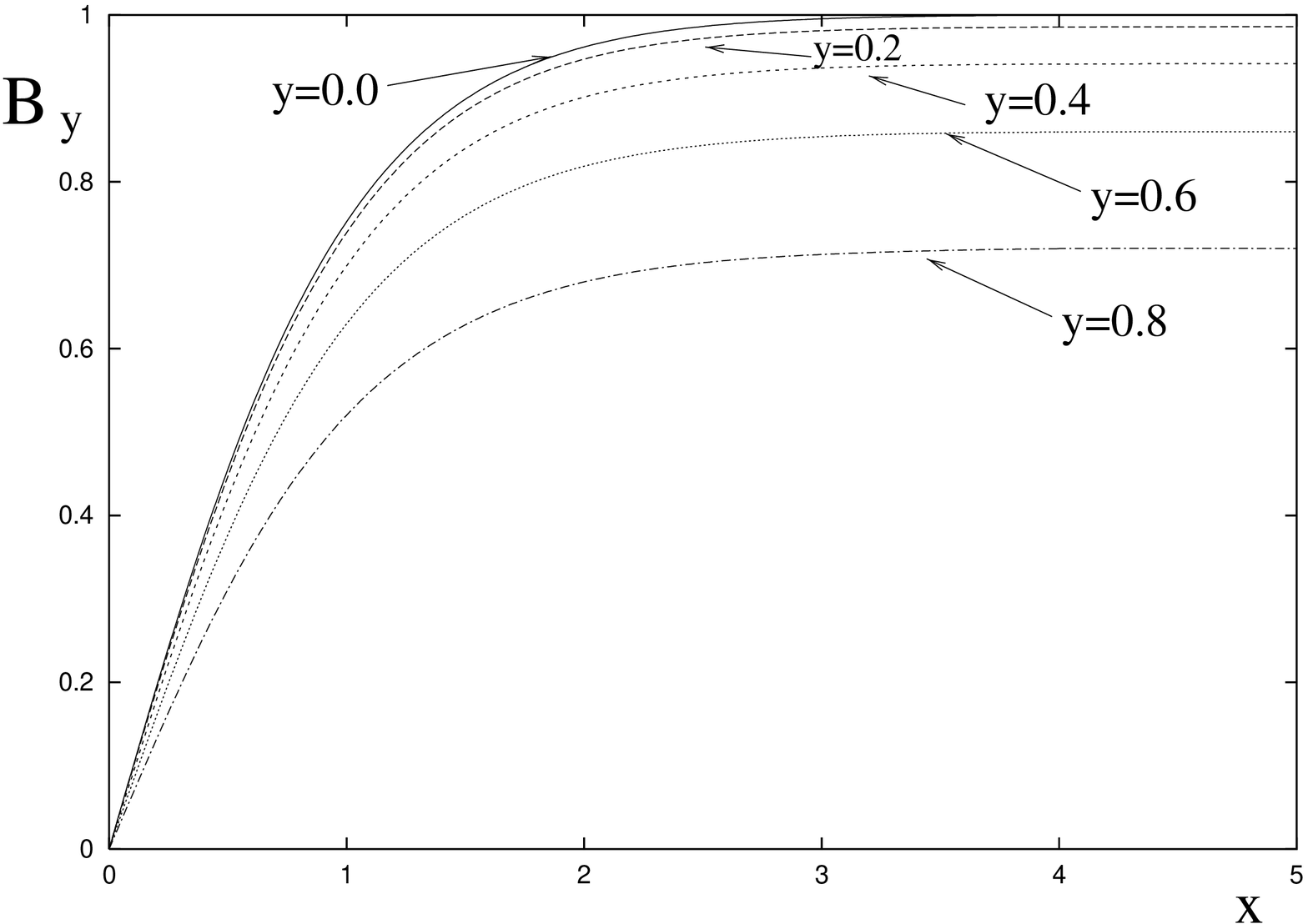,height=3in,width=5in}}
\caption[Plots of $B_y$ as a function of $x$ for several different
values of $y$ in the steady state]
{Plots of $B_y$ as a function of $x$ for several different values of 
$y$ in the steady state.}
\label{fig-local-steadystate-By(x)}
\end{figure}

\begin{figure} [tbp]
\centerline
{\psfig{file=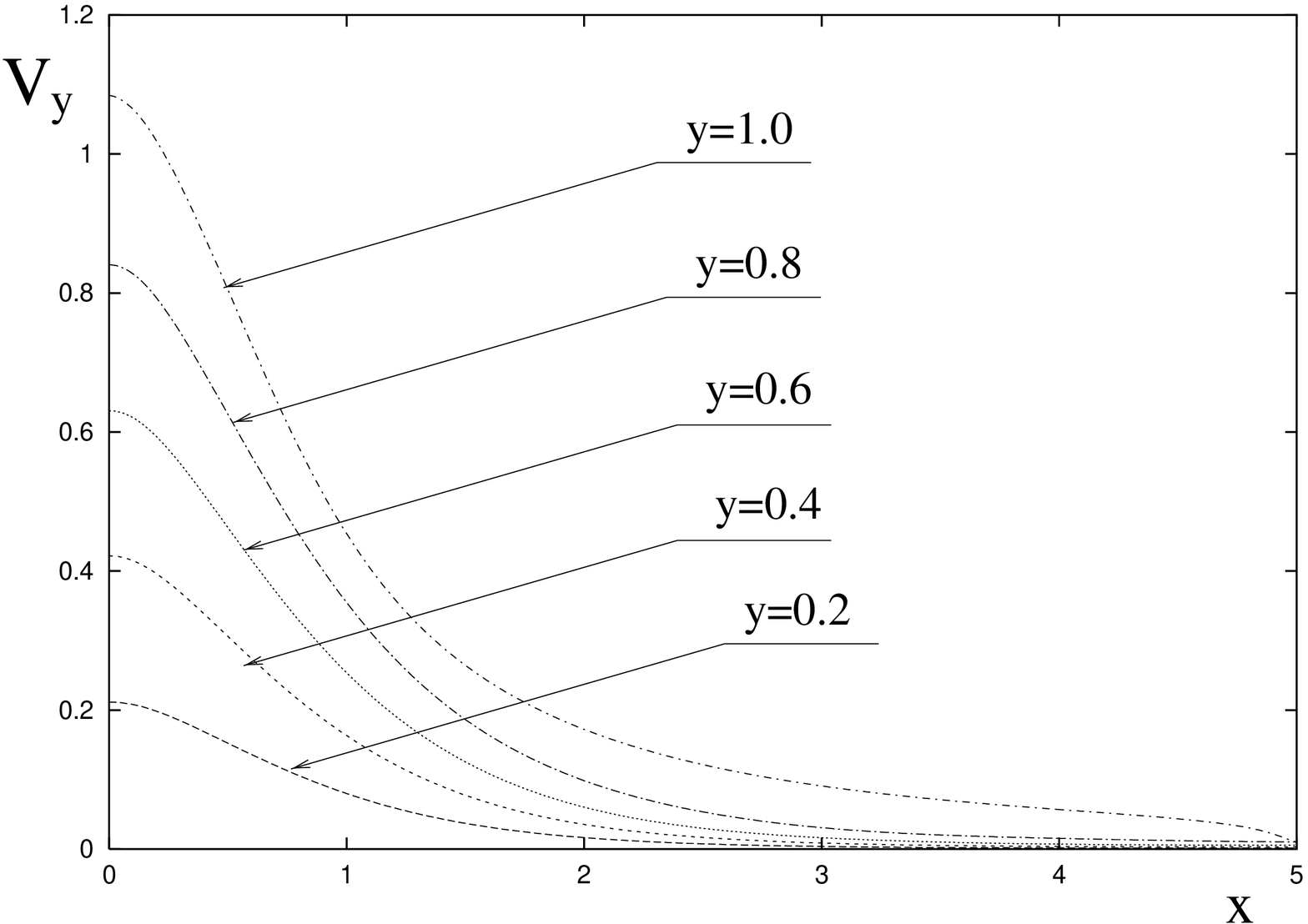,height=3in,width=5in}}
\caption[Plots of $V_y$ as a function of $x$ for several different
values of $y$ in the steady state]
{Plots of $V_y$ as a function of $x$ for several different values of $y$ 
in the steady state.}
\label{fig-local-steadystate-Vy(x)}
\end{figure}

The steady-state Sweet--Parker solution is represented in Figs.~\ref
{fig-local-steadystate-Psi-cont}, \ref{fig-local-steadystate-j-3d},
\ref{fig-local-steadystate-j(x)}, \ref{fig-local-steadystate-By(x)}, 
and \ref{fig-local-steadystate-Vy(x)}. 
This solution corresponds to the following set of the parameters describing 
the boundary conditions:  $x_{\rm lim}=5.0$, $y_{\rm lim}=1.0$, $B_0=0.3$, 
$L_s=2.0$, $C=0.957$ [see Eqs.~(14) and~(29)]. In this particular run
the values of the perpendicular viscosity $\nu$, and of the (artificial) 
resistivity and viscosity acting in the $y$~direction were $\nu=0.02$, 
$\eta_y=\nu_y=0.01$.

There are several things that should be noted about this solution:

As can be seen from Fig.~\ref{fig-local-steadystate-j(x)}, 
the current density as a function of $x$ at any given value 
of $y$ peaks on the midplane $x=0$. This makes the solution
qualitatively different from a Petschek-like configuration 
(see Fig.~\ref{fig-local-init-j-x-petschek}) in which the 
current density is concentrated in a shock-like structure 
off the midplane.

At any given $y$, the current density $j(x)$ rapidly goes to zero 
as $x \rightarrow \infty$, and $B_y(x,y) \rightarrow B_{y,0}(y)$ 
monotonically, which means that there is no flux pile up in front 
of the layer. The velocity $v_y$ does not quite go to zero at the 
upper boundary, but its value at $x=x_{\rm lim}$ is small and goes 
to zero as the artificial $\eta_y, \nu_y \rightarrow 0$.

\begin{figure} [tbp]
\centerline
{\psfig{file=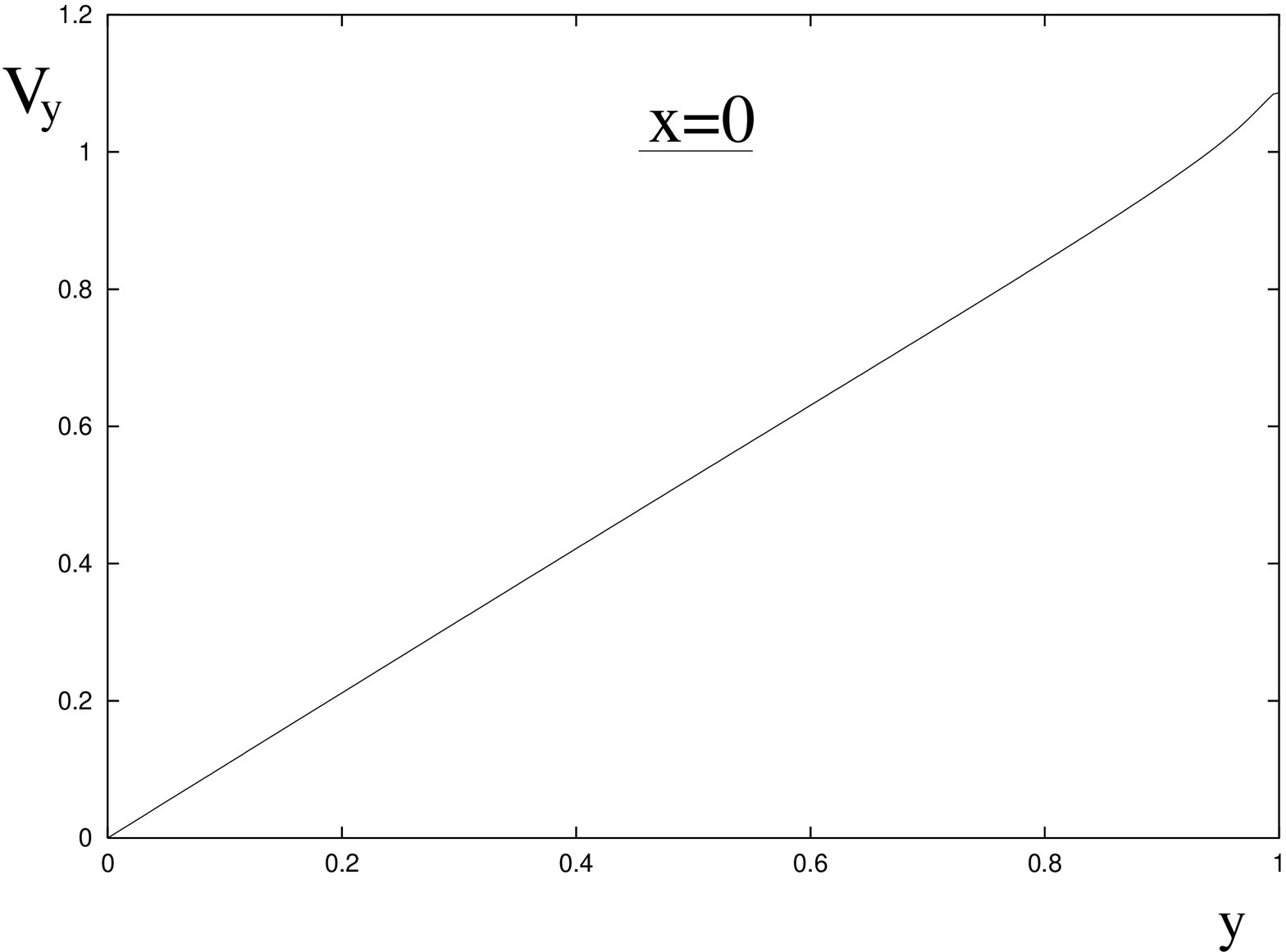,height=3in,width=5in}}
\caption[Plots of $v_y$ as a function of $y$ along the midplane 
$x=0$ in the steady state]
{Plots of $v_y$ as a function of $y$ along the midplane $x=0$ in 
the steady state.}
\label{fig-local-steady-Vy(y)}
\end{figure}

\begin{figure} [tbp]
\centerline
{\psfig{file=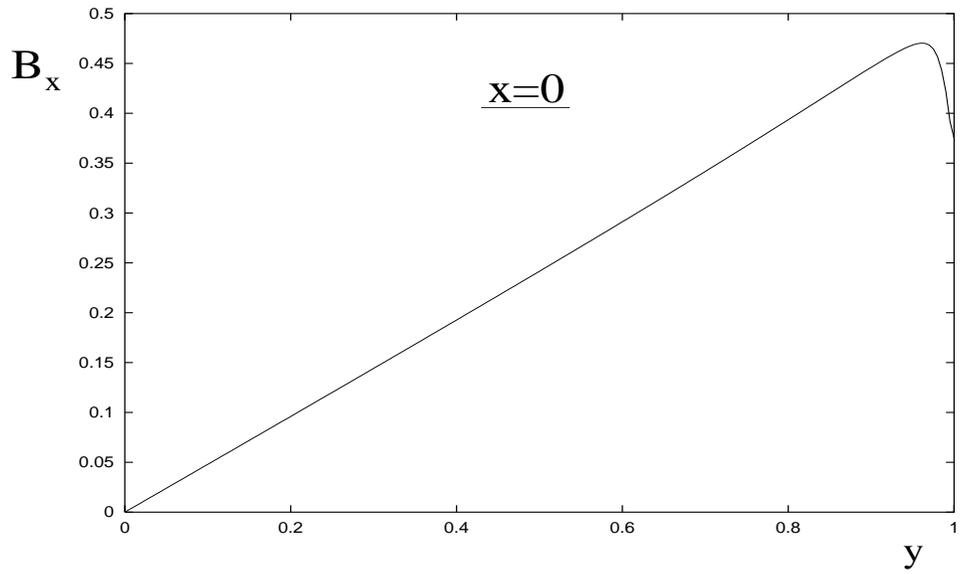,height=3in,width=5in}}
\caption[Plots of $B_x$ as a function of $y$ along the midplane 
$x=0$ in the steady state]
{Plots of $B_x$ as a function of $y$ along the midplane $x=0$ in 
the steady state.}
\label{fig-local-steady-Bx(y)}
\end{figure}

Next, the solution in the layer shows an essentially linear rise of $v_y$ 
and $B_x$ along the midplane $x=0$ (see Figs.~\ref{fig-local-steady-Vy(y)}
and \ref{fig-local-steady-Bx(y)}). The linear behavior of $B_x(x=0,y)$ near 
$y=0$, contrary to the cubic behavior predicted by Priest and Cowley~\cite
{Priest-Cowley-1975}, was explained in Ref.~\cite{Uzdensky-1998} (together 
with the nonanalytic behavior of the solution near the midplane in the 
limit~$\nu\rightarrow 0$).

Finally, $B_x$ exhibit a sharp change near the downstream boundary
$y=y_{\rm lim}$, as can be seen in Figs.~\ref{fig-local-steady-Bx(y)}. 
This change is due to the fact that in the separatrix region we neglect 
the resistive term $\Psi_{xx}$, which is in fact finite. That is, the 
(perpendicular) resistivity effectively has a discontinuity across 
$y=y_{\rm lim}$: $\eta(y<y_{\rm lim})=1$ and $\eta(y>y_{\rm lim})=0$. 
This discontinuity in the equations also shows up in the solution, but 
it is smoothed out over some vicinity of $y_{\rm lim}$ by the artificial 
resistivity and viscosity in the $y$~direction. As these $\eta_y$, $\nu_y$ 
go to zero, the region of the rapid change near $y_{\rm lim}$ becomes 
smaller and smaller.

Let us add a few remarks about the role of the separatrix region 
in our simulations. 

First, it appears that the destruction of the initially-set-up Petschek-like 
configuration and its conversion to the Sweet--Parker-like layer is so robust 
and happens so fast, that it is determined by the dynamics in the main layer 
and by its interaction with the upstream boundary conditions [i.e., with the 
scale of nonuniformity of the outside magnetic field $B_{y,0}(y)$]. 
As a consequence, it has nothing to do with the downstream boundary 
conditions (i.e., with the separatrix region). Therefore, the fact 
that our model of the separatrix does not describe the separatrix 
completely accurately seems to be unimportant, as far as the instability 
of the Petschek solution is concerned.

Thus, we believe that the separatrix region, while providing 
physically reasonable downstream boundary conditions for the 
main layer problem, does not really have a strong effect on 
the principal result that the stable steady-state solution is 
the Sweet--Parker layer with the Sweet--Parker reconnection rate. 
Still, we have to point out that for the solution of the problem 
to be \emph{really} complete, one needs to build a better model 
of the separatrix region. Such a model would include real time 
dependence and resistivity, and also would treat the very near 
vicinity of the endpoint of the reconnection layer. A proper 
consideration of the endpoint region \emph{cannot} be done in 
rescaled variables, and a further rescaling of variables and 
matching is needed.

\section{Conclusions}
\label{sec-conclusions}

To summarize, in this paper we present a definite systematic
solution to a particular clear-cut, mathematically consistent 
problem concerning the {\it internal structure of the reconnection 
layer} within the canonical framework (incompressible 2D MHD with 
uniform resistivity) with the outside field $B_{y,0}(y)$ varying 
on the global scale along the layer. We have first derived a system 
of rescaled equations that should be valid in the limit $S\rightarrow 
\infty$. Then, we have developed a 2D resistive MHD code that followed 
the time evolution of the system in order to achieve the steady state. 

We conclude that, under the assumptions of our model, {\it the Petschek-like 
solutions are unstable} and the system quickly evolves to the only stable 
steady-state solution corresponding to the {\it Sweet--Parker reconnection 
layer}. Thus, the Petschek mechanism for fast reconnection does not work in 
our model. The steady-state reconnection rate in our model problem is 
remarkably close to the Sweet--Parker value $E_{\rm SP}=B_{y,0}(0)V_A/
\sqrt{S}$. 

This main result is consistent with the results of simulations conducted 
by Biskamp\cite{Biskamp-1986} and also those by Ugai\cite{Ugai-Tsuda-1977} 
and by Scholer\cite{Scholer-1989}. It also agrees with the experimental 
results in the MRX experiment \cite{Ji-1998}.

Finally, even though we draw our conclusions (about Petschek-like 
structures being unstable) only for this very specific model, this 
result is fundamentally important, because this model is the canonical 
framework typical of most models of magnetic reconnection, including 
both Sweet--Parker and Petschek. This framework is the simplest possible 
framework for a reconnection problem, and thus provides the necessary 
foundation on top of which one can add more complicated physical processes. 
Because the Sweet--Parker model with the classical (Spitzer) resistivity 
is known to be too slow to explain the very fast time scale for the energy 
release in solar flares, one has to look for physics beyond resistive MHD 
with the Spitzer resistivity. The inclusion of some new physical processes 
into the model (for example, locally enhanced anomalous resistivity is 
probably the most suitable candidate) would  create a very different 
situation in which some Petschek-like structure with fast reconnection 
may be possible.

\section{Acknowledgment}

We are grateful to D.~Biskamp, S.~Cowley, T.~Forbes, M.~Meneguzzi, 
S.~Jardin, M.~Yamada, H.~Ji, S.~Boldyrev, and A.~Schekochihin for 
several fruitful discussions. This work was supported by Charlotte 
Elizabeth Procter Fellowship, by the Department of Energy Contract 
No. DE-AC02-76-CHO-3073, and by NASA's Astrophysical Program under 
Grant NAGW2419.

\section{APPENDIX}
\label{sec-appendix}

We give a rough physical argument for why the length~$L'$ 
of the Petschek diffusion region must be of order the global 
length~$L$, and correspondingly, why the Petschek reconnection 
rate must reduce to the Sweet--Parker rate.

The equation of motion for the $B_x$
field should have two terms: a loss term
$-(v_A /L') B_x$ , and a source term due 
to the nonuniform merging of the incoming
$B_y$ field. An estimate for the size of 
this source term is as follows.  

The boundary condition for the external field 
$B_y$ is determined globally by a field scaled 
to the global length~$L$. For simplicity, we take 
this as 
$$   B_{y,\rm ext} = B_0 (1 - y^2/L^2),                        \eqno{(A1)} $$
and apply it over the length of the Petschek
diffusion layer, $L' \ll L$. 

If the external field were entirely uniform,
then the merging of the field in the diffusive 
region would be uniform and there would be no 
tendency to develop a $B_x$ field. If the boundary 
condition on $B_y$ is satisfied and $L' \ll L$, 
the tendency to develop $B_x$ by nonuniform merging
would be proportional to $L'^2/L^2$, the nonuniformity 
of the external field. There are several terms leading 
to the generation of $B_x$, all of the same order of 
magnitude. 

For example, consider the rate of generation of $B_x$ 
by the resistive diffusive velocity alone. Let the layer 
initially have only a $B_y$ field and be of uniform 
thickness~$\delta$. Then $j \sim B_{y,\rm ext}/\delta$ is 
stronger at $y=0$ than at $y=L'$ by the amount
$B_0/\delta (L'^2/L^2)$, so the electric field
$E=\eta j$ is correspondingly larger. Therefore, 
the rate at which lines enter at $y=0$ is larger by 
$\Delta \dot{\Psi} = (\eta/\delta)(L'^2/L^2) B_0$
than that at $ y= L' $.

These excess lines must turn from the $y$
direction to the $x$ direction and produce
the $B_x$ field. Dividing this rate by $L'$ 
we get the rate of generation of the $B_x$ field,
$$ \dot{B}_x (generation) = (\eta/\delta_0 )(L'/L^2)  B_0.  \eqno{(A2)} $$ 

One can convince oneself that the other terms, that generate 
the $B_x$~field such as the variation in the layer thickness~$\delta$, 
the variation in the $x$ velocity of the plasma, etc., are of the same 
order of magnitude.  Thus, the above rate $(\eta/\delta) (L'^2/L^2) 
(B_0/L')$ should give a rough positive upper bound to the regeneration 
rate of the $B_x$ field. Therefore, the $B_x$  field cannot be stronger 
than that given by the balance equation 
$$ \frac{d B_x}{dt} = - B_x \frac{V_A }{L'} 
     + \frac{\eta }{\delta} \frac{L'}{L^2}B_0 = 0,       \eqno{(A3)} $$
so $B_x= (\eta/\delta) (L'^2/L^2) B_0/V_A $.
Now, according to Petschek, the emerging $B_x$~field 
drives the shocks at a speed $B_x/\sqrt{4\pi\rho}$ in 
the $x$~direction, which just balances the incoming 
plasma with velocity $v_{\rm rec}$.

Thus,   
$$ v_{\rm rec} = \frac{B_x}{\sqrt{4\pi\rho}} =
\frac{\eta }{\delta} \frac{L'^2}{L^2}                     \eqno{(A4)} $$

But we also have, from Petschek,
$ v_{\rm rec} = \eta/\delta_0$, so that
$$ L' = L,                                                 \eqno{(A5)} $$
and this reduces the Petschek reconnection rate
$$v_{\rm rec,Petschek}=\frac{V_A }{\sqrt{S}}\sqrt{\frac{L'}{L}}, \eqno{(A6)} $$
to the Sweet--Parker rate.


\begin{thebibliography}{99}

\bibitem{Kulsrud-1998} R.~M.~Kulsrud, Phys. Plasmas, {\bf 5}, 1599 (1998).

\bibitem{MRX-Yamada} M.~Yamada, H.~Ji, S.~Hsu, T.~Carter, R.~Kulsrud,
Y.~Ono, F.~Perkins, Phys. Rev. Lett., {\bf 78}, 3117 (1997).

\bibitem{Sweet-1958} P.~A.~Sweet, in ``Electromagnetic Phenomena in Cosmical 
Physics'', ed. B.Lehnert, (Cambridge University Press, New York, 1958), p. 123.

\bibitem{Parker-1963} E.~N.~Parker, Astrophysical Journal Supplement Series, 
8, p. 177, 1963.

\bibitem{Petschek-1964} H.~E.~Petschek, AAS-NASA Symposium on Solar Flares, 
(National Aeronautics and Space Administration, Washington, DC, 1964),
NASA SP50, p.425.

\bibitem{Biskamp-1986} D.~Biskamp, Phys. Fluids, {\bf 29}, 1520, (1986).
 
\bibitem{Uzdensky-1996} D.~A.~Uzdensky, R.~M.~Kulsrud, and M.~Yamada, 
Phys. Plasmas, {\bf 3}, 1220, (1996).

\bibitem{Uzdensky-1997} D.~A.~Uzdensky and R.~M.~Kulsrud,
Phys. Plasmas, {\bf 4}, 3960 (1997).

\bibitem{Uzdensky-1998} D.~A.~Uzdensky and R.~M.~Kulsrud,
Phys. Plasmas, {\bf 5}, 3249 (1998).

\bibitem{Thesis} D.~A.~Uzdensky, {\it Theoretical Study 
of Magnetic Reconnection}, Ph.~D.~Thesis, Princeton University,
1998.

\bibitem{Priest-Cowley-1975} E.R. Priest and S.W.H. Cowley, 
J. Plasma Physics, {\bf 14}, part II, 271-282 (1975).

\bibitem{Ugai-Tsuda-1977} M.~Ugai and T.~Tsuda, J. Plasma Phys.,
{\bf 17}, 337 (1977).

\bibitem{Scholer-1989} M.~Scholer, J. Geophys. Res., {\bf 94},
8805 (1989).

\bibitem{Ji-1998} H.~Ji, M.~Yamada, S.~Hsu, R.~Kulsrud, Phys. Rev. Lett.,
{\bf 80}, 3256 (1998).


\end{thebibliography}
\end{document}